\documentclass[twocolumn,twoside,slac_two]{revtex4}
\usepackage{graphicx}
\usepackage{fancyhdr}
\begin{document}

\title{Probing Jet Conditions with Multi-frequency, Centimeter-band Linear Polarization: PKS 0420-01}

%

\author{M. F. Aller, P. A. Hughes, H. D. Aller}
\affiliation{University of Michigan, Astronomy Department, 500 Church St., Ann Arbor, MI, 48109-1042, USA}
\author{T. Hovatta}
\affiliation{Cahill Center for Astronomy \& Astrophysics, California Institute of Technology, 1200 E. California Blvd., Pasadena, CA, 91125, USA}

\begin{abstract}
We have used single-dish centimeter-band, multi-frequency linear polarization and total flux density monitoring observations obtained by UMRAO at 14.5, 8, and 4.8 GHz during $\gamma$-ray flaring to constrain propagating shock-in-jet models; this procedure allows us to probe jet conditions at or near the presumed parsec-scale $\gamma$-ray emission site within the relativistic jet. Results are presented for the blazar 0420-014 during the $\gamma$-ray flare detected by the LAT which peaked in late January 2010. Our work identifies the passage of three forward-moving transverse shocks during the radio-band flaring, a shock Lorentz factor of 12, and an observer's viewing angle of 2$^{\circ}$ and sets limits on the energy distribution of the radiating particles. 
\end{abstract}
\maketitle
\thispagestyle{fancy}

\section{Overview}
To investigate whether shocks are present during $\gamma$-ray flares, and to derive source parameters during $\gamma$-ray flaring we have monitored $\approx$30 blazars for total flux density and linear polarization (hereafter LP) with the University  of Michigan 26-m telescope (UMRAO) at 14.5, 8.0, and 4.8 GHz. Specific goals  of our work are to determine whether evidence for the occurrence of internal shocks can  be found in the LP data; if present, we use comparisons of the UMRAO data with simulations from radiative transfer modeling incorporating propagating shocks  to constrain jet flow conditions during the $\gamma$-ray flaring. We present here results for the HPQ 0420-014, analyzed as part of a pilot study. This source was selected on the basis of showing both strong and resolved radio band flares since 2008.5, and $\gamma$-ray flaring  at GeV energies in the $\it Fermi$-LAT monitoring data.

\section{Historical Character of the Radio-band Variability and Relevance to $\gamma$-ray Flaring}
\begin{figure}
\includegraphics[width=75mm]{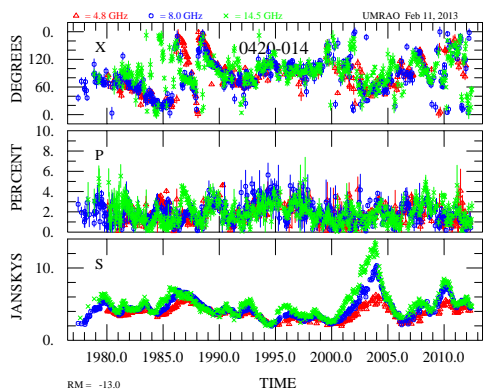}
\caption{ Two-week averages of the long-term UMRAO data showing from bottom to top: S, fractional LP (P\%),  and EVPA. The observations at the three frequencies are symbol and color coded as indicated in the upper left. The source-integrated rotation measure of -13 rad/m$^2$  determined from 18--20.5 cm VLA data \cite{rud83} produces a negligible shift in the UMRAO EVPAs.}
\label{MFAfig1_f1}
\end{figure}

Centimeter-band total flux density (S)  and linear polarization (LP) monitoring observations of the $\gamma$-ray-bright blazar 0420-014 (PKS 0420-01; 2FGLJ0423.2-0120) have been obtained with the University of Michigan (UMRAO) 26-m radio telescope since the late 1970s. These data, as shown in Figure~\ref{MFAfig1_f1}, document that the centimeter-band variability has  been intense and continuous during the past 3.5 decades. A peak amplitude of
S$\approx$14 Jy at 14.5 GHz was reached in late 2003, making this source one of the brightest blazars historically in the centimeter band. Prior to this large outburst, the spectrum was very flat, but the spectrum steepened during the subsequent large flare.  Temporally-resolved flares in fractional LP have also occurred throughout the time period monitored; these observations  identify a  peak amplitude of  5\%. The electric vector position angle (EVPA: shown in the top panel) has exhibited  a series of sharp, systematic changes lasting from several months to years.  Complementary, long-term spatially-resolved VLBI imaging data from the MOJAVE program (at 15.4 GHz) have been obtained since the mid 1990s  with a typical cadence of a few observations per year; these imaging data track the motion of 3 components from 1995 through 2008 and identify a maximum apparent component speed of
$\beta_{app}$=7.36 c (see http:www.physics.purdue.edu/astro/MOJAVE).  In a map obtained in February 2010, a new, very bright component is apparent; this feature is moving at a different position angle from its predecessors and with $\beta_{app}$=3.99$\pm$0.80 c based on 6 epochs of observation \cite{lis13}.

 The source was detected by EGRET in the 1990s, with a highly significant detection (TS=46) in the viewing period centered on 1992.16 \cite{har99} during a plateau in a centimeter-band flare. Shortly after the launch of $\it Fermi$ the source flared in the radio-band, and contemporaneous $\gamma$-ray-radio-band flaring was detected by the LAT. 

An explanation for the cause of the radio-band flaring commonly accepted since the mid-1980s is internal shocks in the parsec-scale region of the jet where the synchrotron emission arises \cite{hug85}, \cite{mar85}. These structures have been shown to develop naturally in hydrodynamic simulations of AGN jets \cite{hug01}, validating this scenario. Several radio band studies based on the development of correlated cross-band activity support the idea that the $\gamma$-ray emission is also produced within the parsec-scale region of the jet, at or near the 43 GHz `core', e.g. \cite{jor12}. Thus, the radio band data probe the region in which the $\gamma$-ray flares are plausibly produced. Evidence for the presence of shocks is of interest since shocks are a potential mechanism for accelerating the emitting particles producing the synchrotron emission to the high energies required for the production of $\gamma$-ray flares. An earlier study of our target by Stevens et al. (\cite{ste95}) compared the spectral evolution of total flux density flares during the period 1989.0-1994, which includes the strong EGRET detection, with the predictions of the Marscher \& Gear \cite{mar85} shock model using high frequency data (375 -- 22 GHz); the spectral behavior was found to be consistent with the shock model predictions. However, discrepancies between the observations and the expected spectral evolution were also identified. These were qualitatively attributed to bending in an initially-straight jet oriented close to the line of sight.

\section{Description of The Model}
In our model, we envisage a propagating shock passing through a medium containing a turbulent magnetic field. As a consequence of the compression associated with the passage of the shock, there is an increase in the fractional linear polarization, and an increase in the particle density (and, hence, the emissivity). The outburst is associated with the propagation of the region bounded by the limits of the shocked flow. In an earlier phase of our program and following the method of \cite{hug89} we developed propagating shock models allowing for the shock to be oriented at an arbitrary angle to the jet flow direction. This work is described in \cite{hug11}. Each shock is specified by its sense (forward or reverse), fractional length relative to the quiescent flow, compression factor, and the shock's obliquity (the orientation relative to the flow direction). The choice of value for an additional free parameter specifying the observer's orientation relative to the shock plane was shown to have little  affect on the light curves. Simulated light curves from radiative transfer calculations based on this scenario may include multiple shocks within a single `event' to explain the variability apparent in the  radio band data. To reduce the number of free parameters, we make the simplifying assumption  that multiple shocks within a single `event' have the same obliquity. A recent refinement of our model has been the incorporation of retarded-time effects. However, comparison of simulated light curves generated with and without retarded-time effects for test cases, shows that only small differences in the simulated flare shapes and spectral behavior result.

\begin{figure}
\centering
\includegraphics[width=85mm]{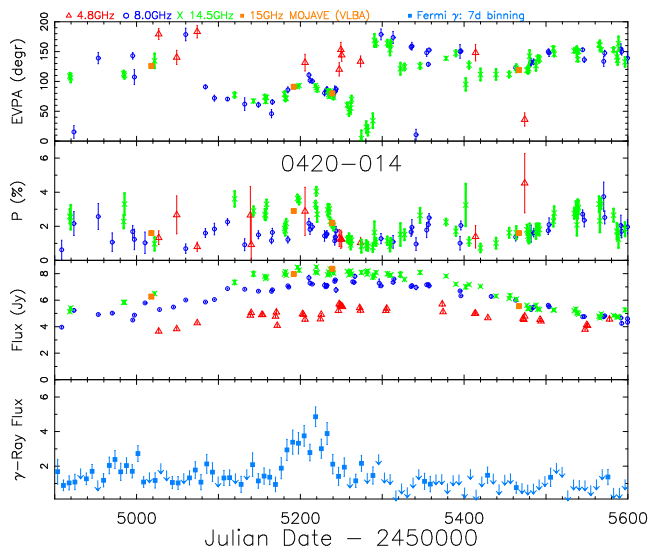}
\caption{The $\gamma$-ray and centimeter band light curves during the radio-band event modeled. The time window shown is March 9, 2009 -- February 7, 2011. From bottom to top: (panel 1) the weekly-binned $\gamma$-ray light curve in units of photons/s/cm$^2$ x 10$^{-7}$,  and (panels 2-4) radio-band total flux density, fractional linear polarization (P\%), and EVPA in the form of daily averages. Source-integrated data from the MOJAVE website are included for comparison.} \label{MFAfig2-f2}
\end{figure}

\section{Comparison of Data and Simulations: Method and Results}
Following the launch of $\it Fermi$ in June 2008, a new phase of high-amplitude, centimeter-band activity commenced. In this paper we study the  2009-2011 radio-band flare contemporaneous with $\gamma$-ray flaring. The combined radio and $\gamma$-ray data during this radio outburst are shown in Figure~\ref{MFAfig2-f2}.  The weekly-binned photon fluxes shown in the bottom panel were obtained using {\bf ScienceTools}--v9r27p1 and P7SOURCE\_V6 event selection. The LAT data were extracted within a 10$^\circ$ region of interest (ROI) centered upon the position of the target. We used an unbinned likelihood analysis (tool gtlike) to determine the photon fluxes by including in the model all of the sources within 15$^{\circ}$ of the target and by freezing the spectral index of all sources to the values in the 2FGL catalogue. For bins with TS$\leq$ 10, we show a 2-$\sigma$ upper limit. A description of the UMRAO observing and reduction procedures is given in \cite{all85}. Several low level $\gamma$-ray flares were detected prior to the main flare which started near RJD 55170 (2009.93). Panels 2-4 show daily averages of the UMRAO total flux density and linear polarization monitoring observations at 14.5, 8.0, and 4.8 GHz. Substructure is clearly apparent in the fractional LP light curve. The centimeter-band LP during the $\gamma$-ray flaring exhibits a systematic swing in the EVPA through 90$^{\circ}$, and an associated increase in the fractional LP; this swing is particularly well-defined by the higher sampling rate at 14.5 GHz. The increase in fractional LP and ordered swing in EVPA are the signature of the passage of a shock through the emitting region.

A serious impediment to a detailed analysis of single dish light curves is the overlapping of multiple, subsequent flares, apparent as structure in the observed light curves. Guided by this structure and using an outburst shape taken from a library of simulations as a template, we estimated the relative times and strengths of each shock to provide a first approximation used  as input for our detailed modeling. Our methodology for the deconvolution of the blended events differs from that adopted by other groups which commonly assumes a generic, exponential flare shape, e.g. \cite{tav11}. 

\begin{figure}
\centering
\includegraphics[width=85mm]{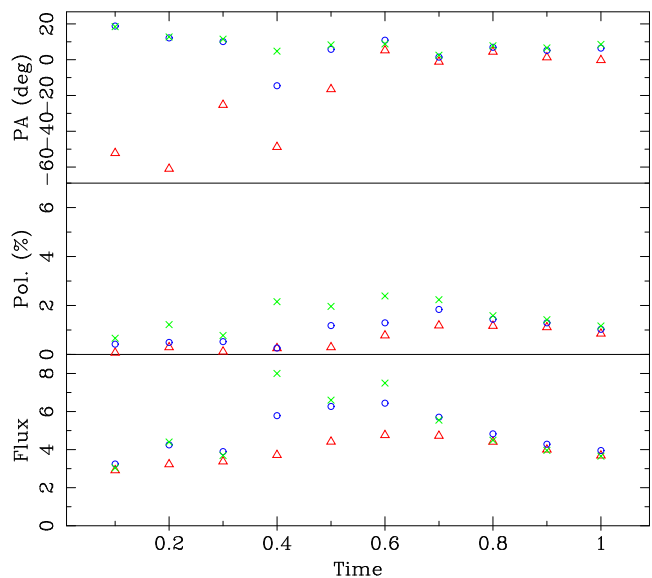}
\caption{Simulation of the radio band light curves during the 2009-2010 event. The computations have been carried out at 3 harmonically-related 
($\sqrt3$) frequencies which correspond to the UMRAO frequencies of 14.5, 8.0, and 4.8 GHz; the symbols follow the convention used in Figures 1 \& 2. The abscissa shows fractional time during the event modeled.} \label{MFAfig3-f3}
\end{figure}
A simulation which reproduces the main features of the data is shown in Figure \ref{MFAfig3-f3}.  These characteristics include the spectral evolution and amplitude of  the total flux density outburst, the magnitude of the change in fractional LP, and the frequency-dependent changes  with time in the EVPA light curves. A comparison of the data and simulations is given in Table \ref{MFA_tab1-t1}. The properties compared are: the range in the amplitude of the total flux density at the highest frequency (14.5 GHz), the evolution in the spectral index, $\alpha_S$, based on the total flux densities at 14.5 and 4.8 GHz, the amplitude of the change (minimum to maximum) of the fractional linear polarization, and the spectral change in the EVPAs during the evolution of the outburst. The latter is complex. As the flare starts, the EVPAs at 4.8 GHz are shifted from those at 14.5 and 8 GHz which track each other, but the EVPAs converge for the three frequencies just beyond the outburst peak. There is a null in the observed P\% and an associated sharp variation in EVPA preceding RJD 55300 which is not reproduced by the model, however. In generating these simulated light curves we incorporated three forward moving shocks oriented transversely to the flow axis, and to characterize the flow we adopted an optically-thin spectral index $\alpha_{thin}$=0.25, a fiducial ``thermal'' Lorentz factor (associated with the random motions of the emitting particles) $\gamma_c$=1000, a low energy cutoff Lorentz factor $\gamma_l$=50, a bulk Lorentz factor of the flow $\gamma_f$=5.0, and an observer's viewing angle of $\theta_{obs}$= 2$^{\circ}$. See \cite{hug11} for definitions and explanations of these parameters. We tabulate in Table \ref{MFA_tab2-t2} the parameters specifying the three shocks. The first shock is weak. The second shock is the main, central one, and it starts at around t=2009.7. The third shock commences around 2010.0, and it is this shock which coincides temporally with the brightest
 $\gamma$-ray flare. This shock is  only 35\% of the length of the second shock and, hence, a new jet component corresponding to this shock might not be expected to be resolved in sequences of VLBI images. However, a new  component is apparent in a MOJAVE imaging map obtained near this epoch as discussed earlier. Based on
the Lorentz factor of the shock of about 12, and a viewing angle of 2$^{\circ}$ we derive a $\beta_{app}$=8 c which is consistent with the MOJAVE results on component speeds. The shocked flow has a Lorentz factor of about 8. The model fitting also allows us to constrain the particle energetics; in a second source recently modeled,
OJ~287, a lower energy cutoff of the `thermal' Lorentz factor was required to obtain a fit to the data.

\begin{table}[t]
\begin{center}
\caption{Comparison of the Observed and Simulated Radio-Band Light Curves}
\begin{tabular}{|l|c|c|c|}
\hline \textbf{Property}  &  \textbf{Observed} & \textbf{Simulated}
\\
\hline Range in S(14.5) &4--8 & 3-- 8 \\
\hline  $\alpha_S$ evolution & flat, inv., flat & flat, inv., flat \\
\hline Range in LP\% & 1-3 &  0-2\\
\hline EVPA behavior & complex  & complex \\
\hline
\end{tabular}
\label{MFA_tab1-t1}
\end{center}
\end{table}

\begin{table}[t]
\begin{center}
\caption{Shock parameters}
\begin{tabular}{|l|c|c|c|}
\hline \textbf{Shock} & \textbf{1} & \textbf{2} & \textbf{3}
\\
\hline obliquity ($\eta$) & 90 & 90 & 90 \\
\hline sense & F &F & F \\
\hline length (l) & 10 & 11 & 4 \\
\hline compression ($\kappa$) & 0.8 & 0.6 & 0.6 \\
\hline
\end{tabular}
\label{MFA_tab2-t2}
\end{center}
\end{table}

\section{Discussion}
Multifrequency, single dish polarimetry data, combined with modeling, have provided sensitive probes of the  jet conditions associated with $\gamma$-ray flaring in this blazar.  For fast flows seen at orientations of only a few degrees, as in this source, small changes in viewing angle make significant differences in the simulated properties, most significantly in the amplitude of the fractional linear polarization. As a consequence, our LP data are a powerful constraint on this important parameter and an independent means for its determination: viewing angle determinations obtained by comparing positions of core and jet features on VLBI maps are sensitive to the segment of the jet dominating the emission at the frequency used for the observations. We note that the goal of this stage of our work has been to validate our assumed scenario, which for the source discussed here incorporates transverse shocks and rectilinear flow. While we wish to explore other fits to this source and to model the variability in other sources in future work, real flows are complex, hydrodynamical entities. Jet curvature in particular  on a variety of scales can affect the observer's interpretation of the observed variability. We caution that the simulations are highly sensitive to the flow detail, so, in reality, it may not be possible to improve dramatically upon the already-good fit described here for 0420-014. The method has yielded results consistent with VLBI determinations while provided more detailed information about the flow conditions where the flaring is produced.
\bigskip 
\begin{acknowledgments}
This work was funded in part by NASA Fermi GI grants NNX09AU16G, NNX10AP16G \& NNX11AO13G. T. Hovatta was supported by a grant from the Jenny and Antti Wihuri foundation. This project has made use of data from the MOJAVE website which is maintained by the MOJAVE team.
\end{acknowledgments}
\bigskip 


\begin{thebibliography}{99} 

\bibitem{lis13} M. Lister, et al., in preparation, 2013.
\bibitem{har99} R.C. Hartman et al., ``The Third EGRET Catalogue of High-Energy Gamma-Ray Sources'', ApJS, 123, 79, 1999.
\bibitem{hug85}  P.A. Hughes, M.F. Aller, \& H.D. Aller, ``Polarized Radio Outbursts in BL-Lacertae. II. The Flux and Polarization of a Piston-Driven Shock'', ApJ, 298, 301, 1985.
\bibitem{mar85} A.P. Marscher \& W.A. Gear, ``Models for High-Frequency Radio Outbursts in Extragalactic Sources with Application to the Early 1983 Millimeter-to-Infrared Flare of 3C~273'', ApJ, 298, 114, 1985.
\bibitem{hug01} P.A. Hughes, M.A. Miller, \& G. C. Duncan, ``Three-Dimensional Hydrodynamic Simulations of Relativistic Extragalactic Jets'', ApJ, 572, 2001.
\bibitem{jor12} S. Jorstad, ``Evidence of a Strong Connection Between Gamma-ray Outbursts and Events in the Millimeter-Wave Core of Blazars'', Presentation at the 4th International Fermi Conference, 2012.
\bibitem{ste95} J.A. Stevens, et al., ``The Spectral Evolution of High-Frequency Radio Outbursts in the Blazar PKS 0420-014'', MNRAS, 275, 1146, 1995.
\bibitem{rud83} L. Rudnick \& T.W. Jones, ``Rotation Measures for Compact Variable Radio Sources'', AJ, 88, 518, 1983.
\bibitem{hug89} P.A. Hughes, H.D. Aller, \& M.F. Aller, ``Synchrotron Emission from Shocked Relativistic Jets. I. The Theory of Radio-Wavelength Variability and its Relation to Superluminal Motion'', ApJ, 341, 54, 1989. 
\bibitem{hug11} P.A. Hughes, M.F. Aller, \& H.D. Aller, ``Oblique Shocks as the Origin of Radio to Gamma-ray Variability in Active Galactic Nuclei'', ApJ, 735, 81, 2011.
\bibitem{all85} H. D. Aller, et al, ``Spectra and Linear Polarizations of Extragalactic Variable Sources at Centimeter Wavelengths'', ApJS, 59, 513, 1985.
\bibitem{tav11} J. Le\'on-Tavares, et al. ``The Connection Between Gamma-Ray Emission and Millimeter Flares in Fermi/LAT blazers'', A\&A, 532, 146, 2011.


\end{thebibliography}
\end{document}